# Magnetic Weyl Semimetal Phase in a Kagomé Crystal

D. F. Liu[1,2]†, A. J. Liang[2,3,4]†, E. K. Liu[5,6]†, Q. N. Xu[5]†, Y. W. Li[7], C. Chen[2,3,7], D. Pei[7], W. J. Shi[2], S. K. Mo[4], P. Dudin[8], T. Kim[8], C. Cacho[8], G. Li[2,3], Y. Sun[5], L. X. Yang[9], Z. K. Liu[2,3], S. S. P. Parkin[1], C. Felser[5,10,11], Y. L. Chen[2,3,7,9]*

[1]*Max Planck Institute of Microstructure Physics, Halle, 06120, Germany*
[2]*School of Physical Science and Technology, ShanghaiTech University, Shanghai, 201210, China*
[3]*ShanghaiTech Laboratory for Topological Physics, Shanghai 200031, P. R. China*
[4]*Advanced Light Source, Lawrence Berkeley National Laboratory, Berkeley, California 94720, USA*
[5]*Max Planck Institute for Chemical Physics of Solids, Dresden, D-01187, Germany*
[6]*Institute of Physics, Chinese Academy of Sciences, Beijing, 100190, China*
[7]*Clarendon Laboratory, Department of Physics, University of Oxford, Oxford OX1 3PU, U.K.*
[8]*Diamond Light Source, Didcot, OX110DE, U.K.*
[9]*State Key Laboratory of Low Dimensional Quantum Physics, Department of Physics and Collaborative Innovation Center of Quantum Matter, Tsinghua University, Beijing, 100084, China*
[10]*John A. Paulson School of Engineering and Applied Sciences, Harvard University, Cambridge, MA 02138, USA*
[11]*Department of Physics, Harvard University, Cambridge, MA 02138, USA*

† These authors contributed equally to this work
* Email: yulin.chen@physics.ox.ac.uk

**Weyl semimetals are crystalline solids that host emergent relativistic Weyl fermions and have characteristic surface Fermi-arcs in their electronic structure. Weyl semimetals with broken time reversal symmetry are difficult to identify unambiguously. In this work, using angle-resolved photoemission spectroscopy, we visualized the electronic structure of the ferromagnetic crystal $Co_3Sn_2S_2$ and discovered its characteristic surface Fermi-arcs and linear bulk band dispersions across the Weyl points. These results establish $Co_3Sn_2S_2$ as a magnetic Weyl semimetal that may serve as a platform for realizing phenomena such as chiral magnetic effects, unusually large anomalous Hall effect and quantum anomalous Hall effect.**

The past decade has witnessed exciting progress in condensed matter physics: relativistic phenomena can be simulated in easily available tabletop materials (*1-5*), and the principles of topology can be used for the discovery of materials with exotic physical properties (*3-5*).

One such class of materials are Weyl semimetals (WSMs), which host emergent Weyl fermions in the bulk and surface Fermi-arc (SFA) states that connect the Weyl points of opposite chirality (*6-11*). This can give rise to unusual physical phenomena (*12-16*), and even inspire theoretical progress (*17-20*). In solids, WSMs can exist in crystals that break the time reversal symmetry (TRS) (*5, 6, 21, 22*), inversion symmetry (IS) (*7-11, 23*), or both (*24*). Compared with the IS-breaking WSMs (*9-11, 25-28*), the TRS-breaking WSMs provide a playground for the interplay between magnetism, electron correlation and topological orders, which can give rise to rich exotic quantum states (Fig. 1A) ranging from quantum anomalous Hall (QAH) effects to Axion insulators (*5, 6, 29-31*).

TRS-breaking WSMs have other preferred properties. For example, chiral anomaly is easier to observe in materials that have only two Weyl points (*5, 32*), which is possible only in TRS-breaking WSMs (Fig. 1B (i)); by contrast, time-reversal invariant IS-breaking WSMs have a minimum of four Weyl points (Fig. 1B (ii)), (*5*)). Additionally, by preserving the IS, the energies of a pair of Weyl points in a TRS-breaking WSM are required to be the same (*5*), making it possible to realize the true nodal WSM phase when the Fermi energy coincides with the Weyl nodes. Finally, the TRS-breaking WSMs are attractive for spintronics applications, as the enhanced Berry curvature, together with its intrinsic magnetism, may lead to unusually large anomalous Hall conductivity (AHC) and the anomalous Hall angle (AHA) (*33, 34*).

However, despite the many proposed candidates (*6, 21, 22, 35-38*), unambiguous and direct experimental confirmation of TRS-breaking WSMs remains challenging.

Recently, a ferromagnetic Shandite $Co_3Sn_2S_2$ was proposed to be a TRS-breaking WSM with three pairs of Weyl points in its 3D Brillouin zone (BZ) (*33, 39*). The transport measurements have demonstrated an unusually large AHC and large AHA (*33, 34*) in this material, making it a promising magnetic WSM candidate. The electronic band structure obtained from theory has shown similarity to experiment (*34*); however, direct evidence for WSMs, such as the existence of bulk Weyl points with linear dispersions, and the SFAs, is still missing. Here we used angle-resolved photoemission spectroscopy (ARPES), to systematically study the electronic structures of single-crystal $Co_3Sn_2S_2$ and observed the characteristic SFAs and the bulk Weyl points in the ferromagnetic phase. These findings, further supported by excellent agreement with ab initio calculations, confirm the TRS-breaking WSM phase in $Co_3Sn_2S_2$, and provide important insights for the understanding of its exotic physical properties (see discussion in (*40*) for details).

The crystal structure of $Co_3Sn_2S_2$ is comprised of stacked … – Sn-[S-($Co_3$-Sn)-S] –… layers (see Fig. 1C, space group $R\bar{3}m$, No. 166). In each [S-($Co_3$-Sn)-S] layer group, the central Co layer forms a 2D Kagomé lattice with an Sn atom at the center of the hexagon; S atoms are located alternately above and below the triangles formed by the Co atoms, with the adjacent [S-($Co_3$-Sn)-S] layer groups linked by layer-sharing Sn atoms (Fig. 1C).

The TRS-breaking WSM phase in $Co_3Sn_2S_2$ (Fig. 1D) is caused by the joint effects of crystal field, ferromagnetism (FM) and the spin-orbital coupling (SOC): the crystal field first mixes the valence band (VB) and conduction band (CB) to form four-fold degenerate nodal lines (black curve in Fig. 1D (ii)); subsequently, the degeneracy of the nodal line is lifted (Fig. 1D (iii), green curve) by the FM transition that breaks the TRS; finally, spin-orbit coupling (SOC) splits the doubly degenerate nodal line in Fig. 1D (iii) into a pair of Weyl points with opposite chirality (Fig. 1D (iv)). According to recent ab initio calculations for this material (*33, 39*), there are three pairs of Weyl points within each bulk BZ connected by the SFAs (Fig.1E).

To study the electronic structure of $Co_3Sn_2S_2$ and its magnetic Weyl semimetal nature, we synthesized high-quality crystals with flat, shiny, cleaved surfaces (*40*). The temperature dependent transport (Fig. 1F) and magnetization measurements (Fig. 1F, inset) clearly illustrate that an FM transition occurs at a critical temperature $T_C$ = 175 K with a hysteresis loop (for additional characterization, see (*40*).

The overall band structure of $Co_3Sn_2S_2$ obtained through ARPES is summarized in Fig. 2. The experimental stacking plots of constant energy contours of the electronic bands at different binding energies (Fig. 2A) show sophisticated structures and their evolution with energy. To understand these rich details, we carried out ab initio calculations (see details in (*40*)) for the bulk electronic bands for comparison. As shown in Fig. 2, B to D, the experimental results and calculations show nice overall agreement, except for the triangle-shape Fermi surface (FS) pieces around the $\bar{K}$ and $\bar{K}'$ points, which were observed in experiments (see Fig. 2C) but absent in the bulk calculations. These unusual FS pieces, as we will demonstrate below, arise from the topological surface states which will result in characteristic SFAs.

After establishing the overall correspondence between the experimental and calculated (bulk) band structures, we now focus on the vicinity of the triangle-shaped FSs by performing fine

ARPES mapping with high resolution to study their detailed geometry and to search for the unusual SFAs, the unique signature of the WSM.

According to our calculations (see Methods section in (*40*) for details), the SFAs in $Co_3Sn_2S_2$ are located around the $\bar{K}'$ of the BZ (Fig. 3A (i), formed by line-segments that connects one pair of Weyl points with opposite chirality in each BZ (Fig. 3A (i)); these line segments from three adjacent BZs can form a triangle-shaped surface FS piece. This unusual surface FS topology was indeed observed experimentally (Fig. 3A (ii-vi)), where the unchanged shape of these line-segment FS pieces from different photon energies indicate their surface origin (Fig. 3A (ii-vi); results from more photon energies can be found in (*40*)). Remarkably, each line segment FS piece merges into the bulk FS pockets near the $\bar{M}'$ point of the BZ (where the Weyl points are located, see Fig. 3A (i)), in excellent agreement with the calculations.

In addition to the FS topology, we can study the dispersions of the topological surface states (TSSs) that result in the SFAs discussed above (see Fig. 3B). The dispersions of the TSSs from different photon energies are in good agreement with calculations (Fig. 3C). Indeed, the photon energy dependent ARPES measurements ( Fig. 3D) show the characteristic vertical dispersionless FS with respect to the photon energy (and thus also $k_z$), unambiguously confirming the surface nature of TSSs (Fig. 3C) and the triangle-shaped FS pieces (in Fig. 3B; more discussion on the surface Fermi-arc states can be found in (*40*)).

With the SFAs identified, we next searched for the characteristic bulk Weyl fermion dispersion. For this purpose, we carried out broad range (50 - 150 eV) photon energy-dependent ARPES measurements (see (*40*) for details) to precisely identify the $k_z$ momentum locations of the Weyl points (Fig. 4A (i)). In Fig. 4A (ii), bulk bands with strong $k_z$ dispersion can be seen in the $k_y$–$k_z$ spectra intensity map (in contrast to the TSSs without $k_z$ dispersion in Fig. 3D), agreeing well with our calculations (overlaid in red on Fig. 4A (ii)). The measured dispersions along two different high-symmetry directions show good overall agreement with calculations (see Fig. 4B; more discussion on the comparison can be found in (*40*)).

The agreement between experiments and calculations in Fig. 4, A and B allows us to identify the bulk Weyl points in $Co_3Sn_2S_2$, which lie at $k_z = \pm 0.086$ Å$^{-1}$ planes (Fig. 4A (i)) and can be accessed by using 115 eV photons (corresponding to $k_z = -0.086$ Å$^{-1}$ in Fig. 4A). To precisely locate the in-plane momentum loci of the Weyl points, we first performed $k_x$–$k_y$ FS mapping (Fig. 4C (i)) of the band structures across the surface BZ, and then we focused on band dispersions that cut through the Weyl point (see the cutting plane in Fig. 4C (i)).

Indeed, the measured bulk band dispersion is linear, matching the calculations (red curve in Fig. 4C (ii)). However, because the Weyl points are located at an energy ~50 meV above the Fermi level ($E_F$) for the undoped sample (see Fig. 4C (ii)), to observe the band crossing at the Weyl point, we introduced in-situ electron doping using an alkaline metal dozer (Fig. 4D, see (*40*) for details) and successfully raised $E_F$ to the Weyl points. As illustrated in Fig. 4E, with the lifting of $E_F$, spot-like FSs (i.e. the Weyl points) emerge (see Fig. 4E (i); at this photon energy, the FS mapping only cuts across three bulk Weyl points at $k_z = -0.086$Å$^{-1}$ plane, as shown in Fig. 4A (i)). The band dispersion in Fig. 4E (ii) also shows the linear crossing of the bands at the Weyl point, in good agreement with calculations.

The observation of the distinctive SFAs and bulk Weyl points with linear dispersions,

together with the overall agreement of the measurements with theoretical calculations, establish $Co_3Sn_2S_2$ as a magnetic WSM. This finding extends the possibilities for the exploration of other exotic phenomena associated with TRS-breaking WSMs (such as the unusually large AHC and QAH effects at the 2D limit) and potential applications. In addition, the topological phase transition across the FM ordering and the detailed spin textures of the SFAs in $Co_3Sn_2S_2$ merit further investigations.

**References and Notes:**

1.  G. E. Volovik, *The Universe in a Helium Droplet* (Oxford University Press, Oxford, 2003).

2.  K. S. Novoselov, *Rev. Mod. Phys.* **83**, 837 (2011).

3.  M. Z. Hasan, C. L. Kane, *Rev. Mod. Phys.* **82**, 3045–3067 (2010).

4.  X.-L. Qi, S.-C. Zhang, *Rev. Mod. Phys.* **83**, 1057–1110 (2011).

5.  N. P. Armitage, E. J. Mele, A. Vishwanath, *Rev. Mod. Phys.* **90**, 015001 (2018).

6.  X. Wan, A. M. Turner, A. Vishwanath, S. Y. Savrasov, *Phys. Rev. B* **83**, 205101 (2011).

7.  H. Weng, C. Fang, Z. Fang, B. A. Bernevig, X. Dai, *Phys. Rev. X* **5**, 011029 (2015).

8.  S.-M. Huang et al., *Nat. Commun.* **6**, 7373 (2015).

9.  S.-Y. Xu et al., *Science* **349**, 613-617 (2015).

10. B. Q. Lv et al., *Phys. Rev. X* **5**, 031013 (2015).

11. L. X. Yang et al., *Nat. Phys.* **11**, 728-732 (2015).

12. A. A. Zyuzin, A. A. Burkov, *Phys. Rev. B* **86**, 115133 (2012).

13. C.-X. Liu, P. Ye, X.-L. Qi, *Phys. Rev. B* **87**, 235306 (2013).

14. K. Landsteiner, *Phys. Rev. B* **89**, 075124 (2014).

15. A. C. Potter, I. Kimchi, A. Vishwanath, *Nat. Commun.* **5**, 5161 (2014).

16. P. Hosur, *Phys. Rev. B* **86**, 195102 (2012).

17. A. Cortijo, Y. Ferreiros, K. Landsteiner, M. A. H. Vozmediano, *Phys. Rev. Lett.* **115**, 177202 (2015).

18. A. G. Grushin, J. W. F. Venderbos, A. Vishwanath, Roni Ilan, *Phys. Rev. X* **6**, 041046 (2016).

19. D. I. Pikulin, A. Chen, M. Franz, *Phys. Rev. X* **6**, 041021 (2016).

20. H. Shapourian, T. L. Hughes, S. Ryu, *Phys. Rev. B* **92**, 165131 (2015).

21. G. Xu, H. Weng, Z. Wang, X. Dai, Z. Fang, *Phys. Rev. Lett.* **107**, 186806 (2011).

22. Z. Wang et al., *Phys. Rev. Lett.* **117**, 236401 (2016).

23. A. A. Soluyanov et al., *Nature* **527**, 495-498 (2015).

24. G. Chang et al., *Phys. Rev. B* **97**, 041104 (R) (2018).

25. Z. K. Liu et al., *Nat. Mater.* **15**, 27-31 (2016).

26. J. Jiang et al., *Nat. Commun.* **8,** 13973 (2017).

27. L. Huang et al. *Nat. Mater.* **15**, 1155-1160 (2016).

28. A. Tamai et al., *Phys. Rev. X* **6**, 031021 (2016).

29. Y. Machida, S. Nakatsuji, S. Onoda, T. Tayama, T. Sakakibara, *Nature* **463,** 210-213 (2010).

30. E. Y. Ma et al., *Science* **350,** 538-541 (2015).

31. T. Kondo et al., *Nat. Commun*. **6,** 10042 (2015).

32. P. Hosur, X.-L. Qi, *C.R. Phys.* **14**, 857 (2013).

33. E. K. Liu et al., *Nat. Phys.* **14,** 1125-1131 (2018).

34. Q. Wang et al., *Nat. Commun.* **9,** 3681 (2018).

35. J. Kübler, C. Felser, *EPL* **114**, 4 (2016).

36. A. A. Burkov, L. Balents, *Phys. Rev. Lett.* **107**, 127205 (2011).

37. K. Kuroda et al., *Nat. Mater.* **16**, 1090-1095 (2017).

38. M. Yao et al., *arXiv*: 1810. 01514.

39. Q. Xu et al., *Phys. Rev. B* **97**, 235416 (2018).

40. See supplementary materials on Science Online.

41. D. Liu, Replication Data for: Magnetic Weyl Semimetal Phase in a Kagomé Crystal, Version 1, Harvard Dataverse (2019).

**Acknowledgments:**

We thank B. H. Yan, for insightful discussions and acknowledge Diamond Light Source beamline I05 (proposal no. SI22367 and SI20683), Advanced Light Source (U.S. DOE Office of Science User Facility under contract no. DE-AC02-05CH11231) BL10.0.1., Stanford Synchrotron Radiation Laboratory (DOE Office of Science User Facility under Contract No. DE-AC02-76SF00515) BL5-2, and Shanghai Synchrotron Radiation Facility beamline 03U (National Natural Science Foundation of China under contract no. 11227902) for access. **Funding:** This work was supported by Shanghai Municipal Science and Technology Major Project (Grant no. 2018SHZDZX02 to Y.L.C., A.J.L., and Z.K.L.), the Alexander von Humboldt Foundation (to D.F.L.), the National Natural Science

Foundation of China (grant nos. 11774190, 11634009, 11674229, and 11974349 to L.X.Y., Y.L.C., Z.K.L., and E.K.L.), the National Key R&D program of China (grant nos. 2017YFA0305400, 2017YFA0206303 to Z.K.L., and E.K.L.), Tsinghua University Initiative Scientific Research Program to Y.L.C., and L.X.Y.), and the Wurzburg-Dresden Cluster of Excellence on Complexity and Topology in Quantum Matter (EXC 2147, no. 39085490 to C.F.). **Author contributions:** Y.L.C. conceived the project. D.F.L. and A.J.L. performed the ARPES experiments with assistance of Y.W.L., C.C., D.P., and E.K.L.. Q.N.X., W.J.S. and Y.S. performed ab initio calculations. E.K.L synthesized and characterized the single crystals. S.K.M., P.D., T.K., C.C. provided the beamline support. L.X.Y., Z.K.L., G.L., S.S.P.P. and C.F. contributed to the scientific discussions. **Competing interests:** The authors declare no competing interests. **Data and materials availability:** All data presented in this paper are available from Harvard Dataverse (*41*).

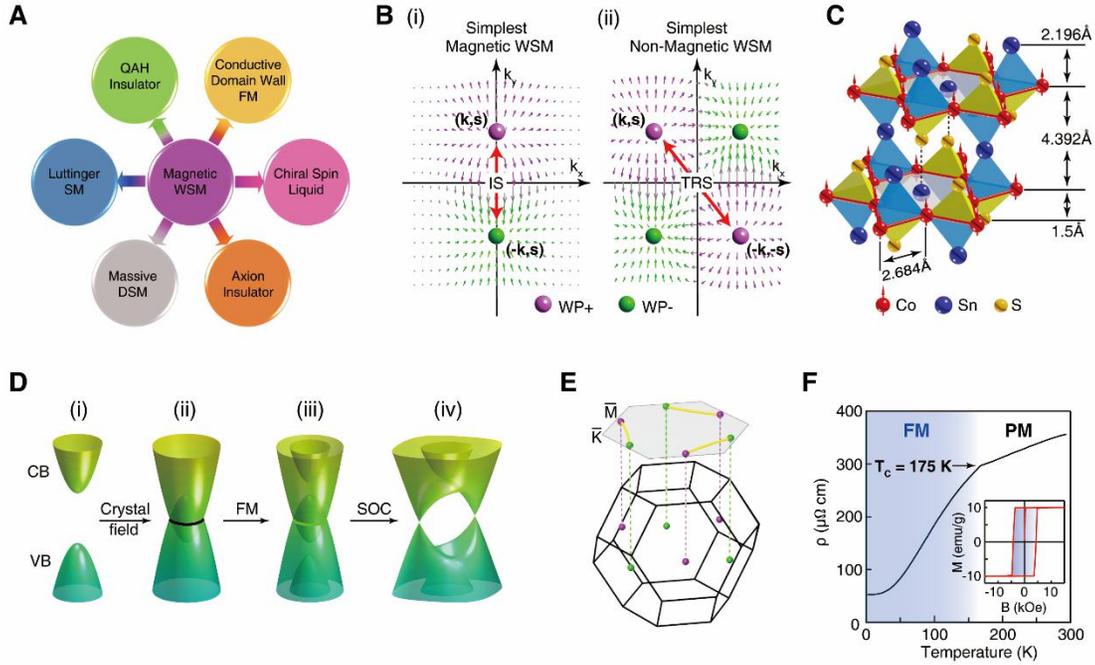

**Figure 1. $Co_3Sn_2S_2$ as a candidate magnetic WSM.**

**(A)** Exotic neighboring states of the magnetic WSM can be achieved by tuning parameters such as magnetism, thickness and electron correlation (see text for more details). The abbreviations stand for: SM, semimetal; DSM, Dirac semimetal; WSM, Weyl semimetal; QAH, quantum anomalous Hall; FM: ferromagnetism. **(B)** Comparison between (i) simplest magnetic WSMs ( with one pair, or two Weyl points) and (ii) non-magnetic WSMs ( with two pairs, or four Weyl points). Magenta and green color of the Weyl points represent positive (+) and negative (-) chirality, respectively; the arrows illustrate the Berry curvature. *k* and *s* stand for momentum and spin, respectively. The abbreviations stand for: WP, Weyl point; IS, inversion symmetry; TRS, time reversal symmetry. **(C)** Crystal structure of $Co_3Sn_2S_2$, showing the stacked …-Sn-[S-($Co_3$-Sn)-S]-… layers. **(D)** The mechanism for magnetic WSM phase in $Co_3Sn_2S_2$, see text for details. The abbreviations stand for: CB, conduction band; VB, valence band; FM, ferromagnetism; SOC, spin-orbital coupling. **(E)** Schematic of the bulk and surface Brillouin zones along the (001) surface of $Co_3Sn_2S_2$, with the Weyl points marked and connected by SFAs (yellow line segments). **(F)** Temperature dependences of longitudinal electric resistivity. The ferromagnetic transition occurs at $T_C$ = 175 K as indicated by the kink in the curve. Inset: Hysteresis loop of the magnetization (external magnetic field is along *z*-axis) measured at T = 2 K, showing a typical ferromagnetic behavior. PM: para-magnetism.

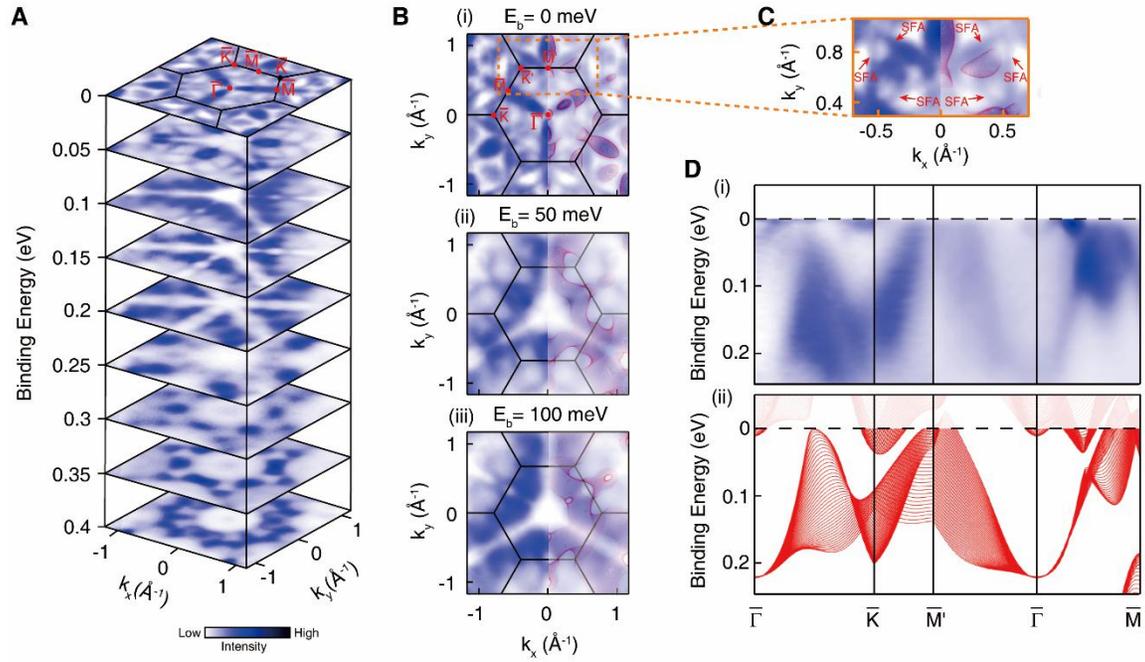

**Figure 2. General electronic structure and band evolution with binding energy.**

**(A)** Stacking plots of constant energy contours at different binding energies obtained from ARPES show sophisticated band structure evolution. **(B)** Comparison between three experimental constant energy contours at different binding energies (i-iii) and the ab initio calculations of the bulk bands (superimposed in red on the right half of the plots), showing excellent agreement except for the triangle shape FSs around the $\overline{K}$ and $\overline{K}'$ points. Note that the experimental plot has been symmetrized according to the crystal symmetry in order to compare with the calculation (as in Fig. 3 and 4). **(C)** Zoomed in Fermi surface around $\overline{K}$ and $\overline{K}'$ points as indicated by the orange dashed line in **B** (i). The triangle shaped FSs around the $\overline{K}$ and $\overline{K}'$ points, which are formed by surface Fermi arcs (SFAs) are marked by red arrows (see text for details). **(D)** Comparison of the experimental (i) and calculated (ii) band dispersions along different high-symmetry directions across the whole BZ

( $\bar{\Gamma} - \bar{K} - \bar{M}' - \bar{\Gamma} - \bar{M}$ ), showing a nice agreement. We note that the calculated bandwidth was renormalized by a factor of 1.43 and the energy position was shifted to match the experiment. The data were recorded at 10 K.

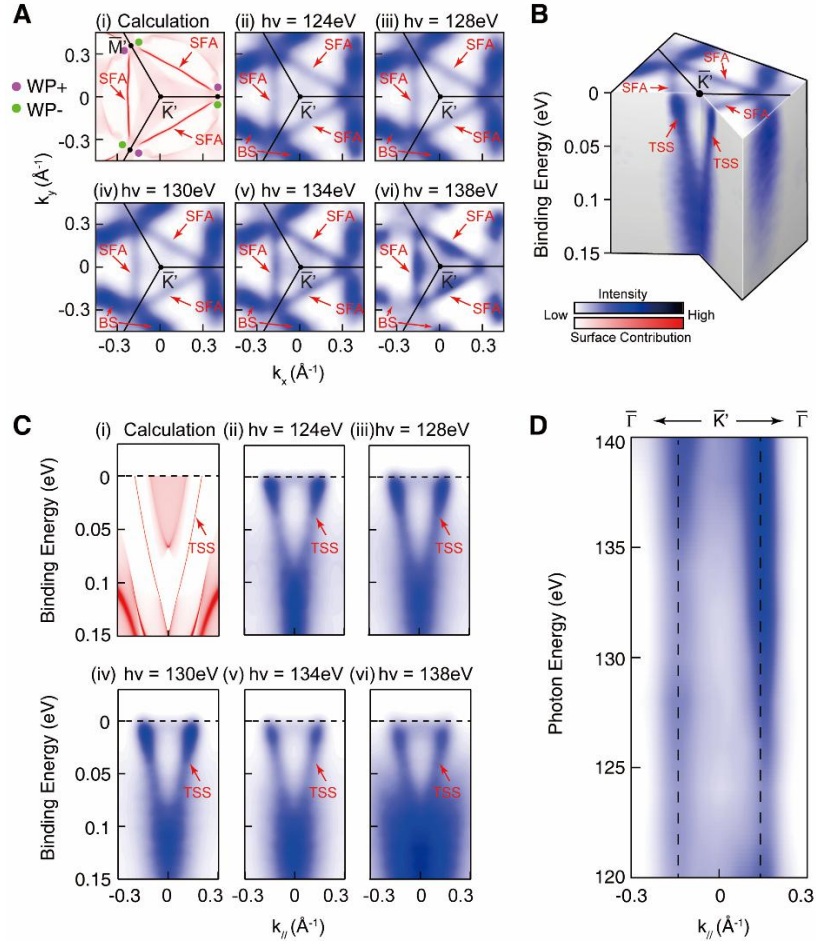

**Figure 3. Observation of the SFAs and TSS dispersion on the (001) surface.**

**(A)** Comparison of the (i) calculated FS from both bulk and surface states (ii-iv) and the experimental FSs under different photon energies. The magenta and green dots in (i) represent the Weyl points with opposite chirality, and the SFAs are indicated by red arrows. (ii)-(vi) Experimental FSs under different photon energies all show SFAs agreeing with the calculation in (i). The triangle-shaped FSs formed by the SFAs do not change in size and shape under different photon energies, confirming their surface origin. **(B)** 3D intensity plot of the experimental band structure near the $\bar{K}'$ point. The SFAs and the dispersions of the topological surface states (TSSs) are marked by red arrows. **(C)** Comparison of the dispersions from (i) calculated TSS along $\bar{\Gamma} - \bar{K}' - \bar{\Gamma}$ direction and (ii)-(vi) the experimental TSSs. **(D)** Photon energy dependent ARPES spectral intensity map at $E_F$ along $\bar{\Gamma} - \bar{K}' - \bar{\Gamma}$ direction, where the black dashed lines indicate the topological surface states that show no dispersion along the energy (and thus $k_z$) direction. The data were recorded at 10 K.

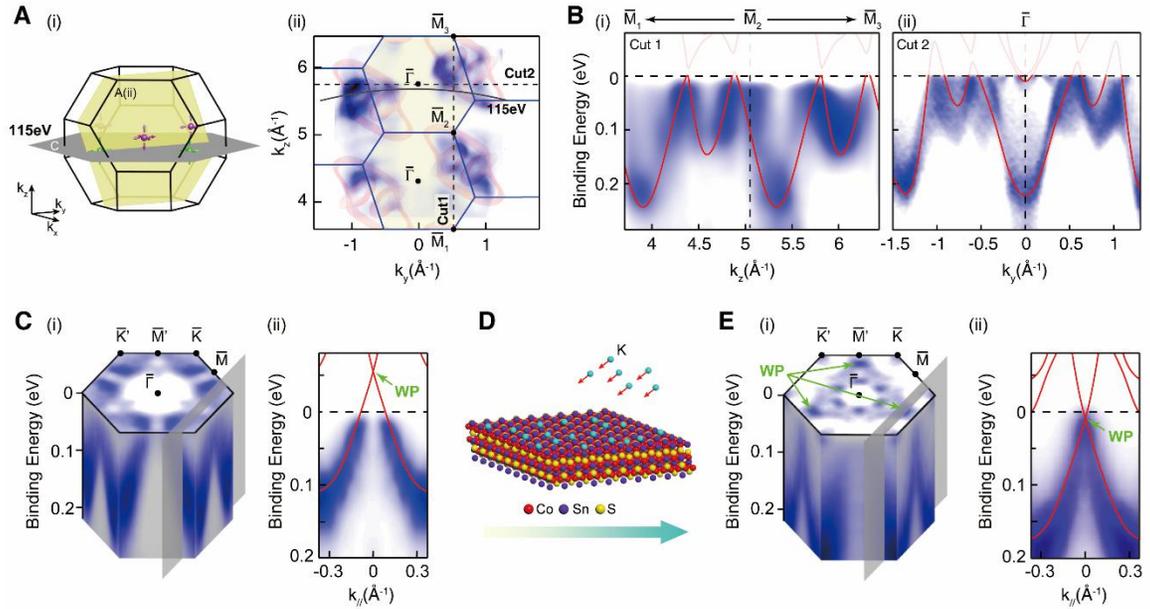

**Figure 4. Bulk band structure and the observation of the Weyl point.**

**(A)** (i) Schematic of the measurement $k_y$-$k_z$ plane (vertical yellow plane) of the intensity plot in (ii). Weyl points are also illustrated, with the reference $k_x$-$k_y$ plane (horizontal gray plane, corresponding to 115eV photons) used to help show their locations. (ii) Photoemission intensity plot along the $k_y$-$k_z$ plane (the yellow plane in (i)). The energy integration window is from $E_F$ - 100 meV to $E_F$. Overlaid red contours are calculated bulk FSs with the same energy integration window, showing overall agreement with the experiment. The black curves indicate the $k_z$ momentum locations probed by 115 eV photons. The two dashed lines marked as 'cut1' and 'cut2' indicate the momentum direction of the two band dispersions shown in **B**. **(B)** Bulk band dispersions along two high-symmetry directions, indicated as cut1 and cut2 in **A** (ii), respectively. The calculated band dispersions (red curves) are overlaid. **(C)** (i) 3D ARPES spectra intensity plot measured with 115 eV photon energy, showing both the FS (top surface) and the band dispersions (side surfaces). The gray plane indicates the location of the band dispersion cut in (ii). (ii) Band dispersion showing linear dispersions toward the Weyl point above the $E_F$, in agreement with the calculations (red curves overlaid). **(D)** Illustration of the in-situ electron doping using an alkaline (potassium) metal dispenser. **(E)** (i) 3D ARPES spectra intensity plot measured after potassium dosing, which lifted $E_F$; Weyl points now emerge (marked by green arrows). (ii) The measured band now shows a linear dispersion across the Weyl point, agreeing well with the calculations (red curves overlaid). The calculated bandwidth was renormalized by a factor of 1.43 and the energy position was shifted to match the experiment. The data were recorded at 10 K.